\documentclass[conference, 10pt]{IEEEtran}

\usepackage{times}
\usepackage[latin1]{inputenc}
\usepackage[T1]{fontenc}
\usepackage[english]{babel}
\usepackage{amsmath, amssymb, graphicx, array, tabularx,booktabs}
\usepackage{epsf,epsfig}
\usepackage{setspace}
\usepackage{subfigure}
\usepackage{algorithmic}
\usepackage{algorithm}
\usepackage{multirow}
\usepackage{url}
\usepackage{color}
\usepackage[nolist,printonlyused]{acronym}
\usepackage{comment}
\usepackage{authblk}
\usepackage{tabularx}
\usepackage{siunitx}
\usepackage{cite}
\usepackage{soul}
\usepackage[font=small,skip=0pt]{caption}
\usepackage{array}
\usepackage{makecell}
\usepackage{enumitem}

\def\q2#1{\textcolor{black}{#1}}

\newcommand{\gf}[1]{\textcolor{black}{#1}}

\newcommand{\ph}[1]{\textcolor{black}{#1}}
\newcommand{\PH}[1]{\textcolor{black}{#1}}
\newcommand{\jv}[1]{\textcolor{black}{#1}}

\newcommand{\fg}[1]{\textcolor{black}{#1}}

\IEEEaftertitletext{\vspace{-1\baselineskip}}

\IEEEoverridecommandlockouts 

\begin{document}
\date{}


\title{Architecture, Protocols, and Algorithms for Location-Aware Services in Beyond 5G Networks}
\singlespacing

\author{
Peter Hammarberg$^{\star}$,
Julia Vinogradova$^{\dag}$,
G\'{a}bor Fodor$^{\star\ddag}$, \\
Ritesh Shreevastav$^{\star}$,
Satyam Dwivedi$^{\star}$,
Fredrik Gunnarsson$^{\star}$,\\
\small $^\star$Ericsson Research, Sweden, E-mail: \texttt{firstname.secondname@ericsson.com} \\
\small $^\dag$Ericsson Research, Finland, Email: \texttt{Julia.Vinogradova@ericsson.com}\\
\small $^\ddag$KTH Royal Institute of Technology, Sweden. E-mail: \texttt{gaborf@kth.se}\\

}


\begin{acronym}[LTE-Advanced]
  \acro{2G}{Second Generation}
  \acro{3G}{3$^\text{rd}$~Generation}
  \acro{3GPP}{3$\text{rd}$~Generation Partnership Project}
  \acro{4G}{4$^\text{th}$~Generation}
  \acro{5G}{5$^\text{th}$~Generation}
  \acro{5GPPP}{5G Infrastructure Public Private Partnership}
  \acro{QAM}{quadrature amplitude modulation}
  \acro{ADAS}{Advanced driver assistance system}
  \acro{AD}{autonomous driving}
  \acro{A-GNSS}{Assisted Global Navigation Satellite System}
  \acro{AI}{artificial intelligence}
  \acro{AMF}{Access and Mobility Management Function}
  \acro{AOA}{angle of arrival}
  \acro{AOD}{angle of departure}
  \acro{API}{application programming interface}
  \acro{AR}{autoregressive}
  \acro{ARQ}{automatic repeat request}
  \acro{ATO}{automatic train operation}
  \acro{BER}{bit error rate}
  \acro{BLER}{block error rate}
  \acro{BPC}{Binary Power Control}
  \acro{BPSK}{Binary Phase-Shift Keying}
  \acro{BRA}{Balanced Random Allocation}
  \acro{BS}{base station}
  \acro{CAM}{cooperative awareness messages}
  \acro{CAP}{Combinatorial Allocation Problem}
  \acro{CAPEX}{capital expenditure}
  \acro{CBF}{coordinated beamforming}
  \acro{CBR}{congestion busy ratio}
  \acro{CDD}{cyclic delay diversity}
  \acro{CDF}{cumulative distribution function}
  \acro{CDL}{clustered delay line}
  \acro{CS}{Coordinated Scheduling}
  \acro{C-ITS}{cooperative intelligent transportation system}
  \acro{CSI}{channel state information}
  \acro{CSIT}{channel state information at the transmitter}
  \acro{D2D}{device-to-device}
  \acro{DCA}{Dynamic Channel Allocation}
  \acro{DCI}{downlink control information}
  \acro{DE}{Differential Evolution}
  \acro{DENM}{decentralized environmental notification messages}
  \acro{DFO}{Doppler frequency offset}
  \acro{DFT}{Discrete Fourier Transform}
  \acro{DIST}{Distance}
  \acro{DL}{downlink}
  \acro{DL-AOD}{Downlink angle of departure}
  \acro{DL-PRS}{downlink positioning reference signal}
  \acro{DL-TDOA}{Downlink time difference of arrival}
  \acro{DMA}{Double Moving Average}
  \acro{DMRS}{Demodulation Reference Signal}
  \acro{D2DM}{D2D Mode}
  \acro{DMS}{D2D Mode Selection}
  \acro{DMRS}{demodulation reference symbol}
  \acro{DPC}{Dirty paper coding}
  \acro{DPS}{Dynamic point switching}
  \acro{DRA}{Dynamic resource assignment}
  \acro{DSA}{Dynamic spectrum access}
  \acro{eMBB}{enhanced mobile broadband}
  \acro{eV2X}{Enhanced vehicle-to-everything}
  \acro{ECID}{Enhanced cell identity}
  \acro{EIRP}{equivalent isotropically radiated power}
  \acro{ERTMS}{European Rail Traffic Management System}
  \acro{ETCS}{European Traffic Control System}
  \acro{ETSI}{European Telecommunications Standards Institute}
  \acro{FDD}{frequency division duplexing}
  \acro{FR1}{frequency range 1}
  \acro{FR2}{frequency range 2}
  \acro{FRMCS}{Future Rail Mobile Communications System}
  \acro{GNSS}{global navigation satellite system}
  \acro{GoA}{grade of automation}
  \acro{GSM-R}{Global System for Mobile Communications for Rail}
  \acro{HARQ}{hybrid automatic repeat request}
  \acro{HST}{high-speed train}
  \acro{IAB}{integrated access and backhaul}
  \acro{ITS}{Intelligent transportation system}
  \acro{KPI}{key performance indicator}
  \acro{IEEE}{Institute of Electronics and Electrical Engineers}
  \acro{IMT}{International Mobile Telecommunications}
  \acro{IMU}{inertial measurement unit}
  \acro{InC}{in-coverage}
  \acro{IoT}{Internet of Things}
  \acro{ISD}{inter-site distance}
  \acro{LDPC}{low-density parity-check coding}
  \acro{LMF}{Location Management Function}
  \acro{LMR}{land mobile radio}
  \acro{LoS}{line-of-sight}
  \acro{LTE}{Long Term Evolution}
  \acro{MAC}{medium access control}
  \acro{mmWave}{millimeter-wave}
  \acro{MBB}{mobile broadband}
  \acro{MCS}{modulation and coding scheme}
  \acro{MEC}{mobile edge computing}
  \acro{METIS}{Mobile Enablers for the Twenty-Twenty Information Society}
  \acro{MIMO}{multiple-input multiple-output}
  \acro{MISO}{multiple-input single-output}
  \acro{ML}{machine learning}
  \acro{MRC}{maximum ratio combining}
  \acro{MS}{mode selection}
  \acro{MSE}{mean square error}
  \acro{MTC}{machine type communications}
  \acro{multi-RTT}{Multi-cell round trip time}
  \acro{multi-TRP}{multiple transmission and reception points}
  \acro{mMTC}{massive machine type communications}
  \acro{cMTC}{critical machine type communications}
  \acro{NDAF}{Network Data Analytics Function}
  \acro{NF}{network function}
  \acro{NLoS}{non-line-of-sight}
  \acro{NR}{New Radio}
  \acro{NSPS}{national security and public safety}
  \acro{NWC}{network coding}
  \acro{OEM}{original equipment manufacturer}
  \acro{OFDM}{orthogonal frequency division multiplexing}
  \acro{OoC}{out-of-coverage}
  \acro{PRS}{positioning reference signal}
  \acro{PSBCH}{physical sidelink broadcast channel}
  \acro{PSFCH}{physical sidelink feedback channel}
  \acro{PSCCH}{physical sidelink control channel}
  \acro{PSSCH}{physical sidelink shared channel}
  \acro{PDCCH}{physical downlink control channel}
  \acro{PDCP}{packet data convergence protocol}
  \acro{PHY}{physical}
  \acro{PLNC}{physical layer network coding}
  \acro{PPPP}{proximity services per packet priority}
  \acro{PPPR}{proximity services per packet reliability}
  \acro{PSD}{power spectral density}
  \acro{RLC}{radio link control}
  \acro{QAM}{quadrature amplitude modulation}
  \acro{QCL}{quasi co-location}
  \acro{QoS}{quality of service}
  \acro{QPSK}{quadrature phase shift keying}
  \acro{PaC}{partial coverage}
  \acro{RAISES}{Reallocation-based Assignment for Improved Spectral Efficiency and Satisfaction}
  \acro{RAN}{radio access network}
  \acro{RA}{Resource Allocation}
  \acro{RAT}{Radio Access Technology}
  \acro{RB}{resource block}
  \acro{RF}{radio frequency}
  \acro{RRC}{Radio Resource Control}
  \acro{RS}{reference signal}
  \acro{RSRP}{Reference Signal Received Power}
  \acro{RTK}{real-time kinematic}
  \acro{SA}{scheduling assignment}
  \acro{SFN}{Single frequency network}
  \acro{SNR}{signal-to-noise ratio}
  \acro{SINR}{signal-to-interference-plus-noise ratio}
  \acro{SC-FDM}{single carrier frequency division modulation}
  \acro{SFBC}{space-frequency block coding}
  \acro{SCI}{sidelink control information}
  \acro{SL}{sidelink}
  \acro{SLAM}{simultaneous localization and mapping}
  \acro{SPS}{semi-persistent scheduling}
  \acro{STC}{space-time coding}
  \acro{SW}{software}
  \acro{TCI}{transmission configuration indication}
  \acro{TBS}{transmission block size}
  \acro{TEG}{timing error group}
  \acro{TDD}{time division duplexing}
  \acro{TOA}{time of arrival}
  \acro{TRP}{transmission and reception point}
  \acro{TTI}{transmission time interval}
  \acro{UAV}{unmanned aerial vehicle}
  \acro{UAM}{urban air mobility}
  \acro{UE}{user equipment}
  \acro{UL}{uplink}
  \acro{UL-AOA}{Uplink angle of arrival}
  \acro{UL-SRS}{Uplink sounding reference signal}
  \acro{UL-TDOA}{Uplink time difference of arrival}
  \acro{URLLC}{ultra-reliable and low latency communications}
  \acro{VUE}{vehicular user equipment}
  \acro{V2I}{vehicle-to-infrastructure}
  \acro{V2N}{vehicle-to-network}
  \acro{V2X}{vehicle-to-everything}
  \acro{V2V}{vehicle-to-vehicle}
  \acro{V2P}{vehicle-to-pedestrian}
  \acro{ZF}{Zero-Forcing}
  \acro{ZMCSCG}{Zero Mean Circularly Symmetric Complex Gaussian}
 \acro{TBS}{transport block size}
 \acro{SCI}{sidelink control information}
 
 \acro{UEA}{\ac{UE} assisted}
 \acro{UEB}{UE based}
\end{acronym}

\maketitle
\pagestyle{empty}
\thispagestyle{empty}

\begin{abstract}
\PH{The automotive and railway industries are rapidly transforming with a strong drive towards automation and digitalization, with the goal of increased convenience, safety, efficiency, and sustainability.}
\PH{Since} assisted and fully automated automotive and train transport services \gf{increasingly rely} on vehicle-to-everything communications, and high-accuracy real-time positioning, it is necessary to continuously maintain high-accuracy localization, even in occlusion scenes such as tunnels, urban canyons, or areas covered by dense foliage. In this paper, we review the 5G positioning framework of the 3rd Generation Partnership Project in terms of methods and architecture and propose enhancements to meet the stringent requirements imposed by the transport industry. In particular, we highlight the benefit of fusing cellular and sensor measurements and discuss required architecture and protocol support for achieving this at the network side. We also propose a positioning framework to fuse cellular network measurements with measurements by onboard sensors. We illustrate the viability of the proposed fusion-based positioning approach using a numerical example.
\\

\end{abstract}

\section{Introduction} 
\label{Sec:Intro}

Transportation systems in the road, rail and aerial transport segments increasingly employ
coordination, automation, electrification and \ac{AI} to enhance functional safety, efficiency and sustainability \cite{Zeadally:20}.
These systems, \gf{which are} often referred to as \acp{C-ITS}, 
rely on information about time and location of objects and events in the surrounding environment \cite{Siegel:18}.
To enable this, communication between
vehicles, vulnerable road users (\gf{such as} pedestrians), other
vehicles and the cellular infrastructure -- that is \ac{V2X} communications -- are instrumental.

Examples of \ac{C-ITS} services include high-definition sensor sharing,
vulnerable road user collision warning, cooperative \gf{maneuvers} of autonomous vehicles for
emergency situations, high-definition map collecting and sharing, and supporting tele-operated driving.
These applications require position information
in real-time with decimeter-level accuracy \cite{5GAA:21}.
Similarly, use cases in urban rail and \ac{HST} scenarios,
such as the unattended rail operations use case,
require positioning at an accuracy beyond \gf{that provided by} current state-of-the-art positioning
schemes.

The \ac{5G} \PH{wireless cellular network} is
designed with many industrial application requirements in mind, supporting large signal bandwidths, very
high data rates, multiple antennas, latencies in the order of 1 ms, and flexibility in terms of
network architecture, carrier frequencies and deployment options \cite{Parkvall:20}.
\ac{5G} systems
aim \gf{to enable} a wide range of positioning capabilities
to meet the requirements from different verticals including automotive and rail transport \cite{TS:22.261}.

Along a related research line, several recent works have proposed positioning methods that are
deployable in \ac{5G} networks, \PH{building} on the signal characteristics of \PH{the \ac{3GPP} \ac{NR} standard}, and serve as key enablers
of real-time localization services in \acp{C-ITS}, including road, rail, and aerial transport \cite{5GAA:21}, \cite{Wymeersch:17}, \cite{Talvitie:19}.
One of the insights that these papers provide is that existing \gf{\acp{GNSS}} alone cannot provide reliable accurate positioning information in  urban areas with tall buildings or in areas with dense foliage. However, for relative positioning, onboard sensors such as accelerometers, gyroscopes, cameras, radars and lidars can operate well, whereas cellular networks can provide absolute positioning. Specifically, it was shown in
\cite{Wen:21}, 
that \ac{mmWave} signals and large \ac{MIMO} \PH{antenna} deployments \gf{enable} technologies for accurate
positioning and device orientation estimation even with only one \ac{BS}.
Therefore, fusing sensory data provided by onboard sensors with radio
access network measurements is an intuitively appealing approach to positioning in \ac{GNSS}-problematic
areas. \gf{A framework} that is based on combining sensory data with cellular signals is referred to sensor
fusion.

Several related works have proposed \gf{using} statistical signal processing techniques to fuse information
from several sensors mounted on vehicles. Some of these schemes are based on a simple odometric model of the
vehicle and a model of each sensor relating to the vehicle \cite{Karlsson:17}. It can be argued that when
road map information -- which cannot be approximated with a Gaussian model -- is utilized, particle filters are
\gf{more advantageous than} Kalman filter algorithms. On the other hand, when the time evolution of angular
information in cellular signals and the movement of the vehicle can be described as an \ac{AR} process,
Kalman filter and extended Kalman filter-based approaches can be employed to combine sensory data with cellular measurements \cite{Mostafavi:20}. While the results reported in the above research papers are encouraging, some of the use cases in advanced \ac{C-ITS} scenarios demand real-time positioning of accuracy well below the meter or even decimeter level (see Table~I) \cite{5GAA:21}.

In this paper we argue that \gf{by building on} and improving the evolving capabilities of \ac{5G} wireless networks, it is possible to reach sub-meter positioning accuracy.
\gf{This requires} not only enhancement of the existing 5G positioning methods,
but also appropriate architecture and protocol support for fusing onboard \gf{sensor measurements and cellular signals}\PH{, specifically in scenarios with large vehicular density.}

The rest of this paper is structured as follows. The next section reviews some of the most important
location aware services and associated requirements in the transport sector. \gf{Section \ref{Sec:5GEvo}}
discusses the evolution of positioning capabilities of \ac{5G} networks.
Section \ref{Sec:Arch} discusses
architecture alternatives that facilitate sensor fusion for location aware transport services.
\gf{Section \ref{Sec:Perf}} presents numerical results.
Section \ref{Sec:Conc} summarizes the paper
and discusses open research challenges.

\section{\gf{Location-Aware} Services and Requirements for Road and Railway Transport} 
\label{Sec:Serv}

\subsection{Service Requirements for Road Transport}

For road transport, \ac{V2X} communication is a key enabling technology for advanced \ac{C-ITS} services \cite{5GAA:21}.
These services collectively aim to improve driver and
passenger convenience, ensure safety of road users and make road transport much more efficient.
As one of the key enablers of \ac{V2X} services, \gf{highly accurate} and up-to-date positioning information
is an indispensable component (see Table \ref{tab:tab1}). \ac{V2X} high-accuracy positioning is also the basic premise
for future \ac{V2X} services such as automated and remote driving \cite{5GAA:21}.

Some of the \acp{KPI} characterizing the quality of positioning information have been extensively discussed in
standardization fora, industrial associations and multinational projects; see, \gf{for example,} \cite{5GAA:21}.
These \acp{KPI}
include positioning accuracy, latency, update rate, and reliability.
Specifically, for \ac{V2X} scenarios, some
other positioning characteristics, such as continuity, security/privacy and cost are also used to characterize
and compare the advantages and disadvantages of positioning solutions.

Table \ref{tab:tab1} lists some \ac{V2X} services and associated service characteristics in terms of vehicle velocity, vehicle density, and positioning accuracy requirements with associated confidence interval levels expressed as their $\sigma$-values (3-$\sigma$ corresponding to the 99.7 percentile confidence interval).
The accuracy requirements range from \gf{tenths} of meters down to sub-meters.
The strictest requirements are seen in areas related to autonomous drive and advanced driver assist features.
For high-definition map collecting and sharing, cooperative \gf{maneuvering}, and tele-operated driving, {accuracies down to 10 cm are required, while 1.5 m suffices for intersection movement assist and lane change warning services.}

\begin{table}[H]
	\centering
	\caption{Some V2X use cases and required positioning indicators {\it (Source: 5G Automotive Association, 5GAA
			and Satellite Technology for Advanced Railway Signalling Project \cite{STARS:17}.)}}
	\vspace{2mm}
	\label{tab:tab1}
	\footnotesize
	\begin{tabular}{
			|m{0.14\textwidth}|
			>{\centering}m{0.08\textwidth}|
			>{\centering}m{0.06\textwidth}|
			m{0.1\textwidth}|
		}
			\hline
			\hline
			\textbf{~~Use case} & \textbf{Velocity [km/h]} & \textbf{Vehicle density $\left[\frac{1}{\text{km}^2}\right]$} & \textbf{\thead{Positioning \\ accuracy [m]}}
			\vspace{1mm} \\
			\hline
			\hline
			Intersection movement assist & 120 & 12000 & 1.5 $(3\sigma)$  \\
			\hline
			Traffic jam warning (urban environment) & 70 & 12000 & 20 $(1\sigma)$ \\
			\hline
			Lane change warning & Host vehicle: 40; Remote vehicle: 50 & 12000 & 1.5 $(3\sigma)$ \\
			\hline
			High-definition sensor sharing & 250 & 12000 & 0.1 $(3\sigma)$ \\
			\hline
			Vulnerable road user (VRU) awareness -- potentially dangerous situation & Urban: 70; \\ Rural:120 & VRU: 300; Vehicles: 1500 & 1 $(3\sigma)$ \\
			\hline
			Real-time situational awareness and high-definition maps & 250 & 1500 & 0.5 $(3\sigma)$ \\
			\hline
			Group start & 70 & 3200 & 0.2 $(3\sigma)$ \\
			\hline
			Tele-operated driving support & 10 & 10 & 0.1 $(3\sigma)$ \\
			\hline
			High-definition map collecting and sharing & City: 70 \\ Highway: 250 & 12000 & 0.1-0.5 $(3\sigma)$ \\
			\hline
			Automated intersection crossing & Urban: 70 \\ Rural: 120 & 3200 vehicles \\ 10000 VRUs & 0.15 $(3\sigma)$ \\
			\hline
			Infrastructure assisted environment perception  & 250 & 1200 & 0.15 $(3\sigma)$ \\
			\hline
			\hline
			Driverless train  & 150 & N/A & 0.25 \\
			\hline
			Location aware beamforming for HST & 500 & N/A & (not applicable)  \\
			\hline
		\end{tabular}
	\end{table}

\subsection{Service Requirements for Rail Transport}
The rail ecosystem is currently transitioning \PH{towards} a fully digitalized, connected, and automated
transport system.
The foundation of this digitalization is the \ac{FRMCS}, driven by the International Union of Railways (UIC).
While \ac{FRMCS} will ultimately replace the legacy rail communications and control services
based on the legacy \ac{GSM-R} system, \ac{FRMCS} goes beyond being a new technology running over
3GPP communications networks.
\PH{Instead,} \ac{FRMCS} is designed to be \gf{bearer- and
radio-technology} independent, allowing a growing set of \ac{C-ITS} services to take advantage of new
features of the underlying communications technologies \cite{Talvitie:19},
\cite{STARS:17}. 
In Europe, for example, \ac{FRMCS} will gradually take over the role of \ac{GSM-R}
as a key enabler of the \ac{ETCS}, which is part of the \ac{ERTMS}, whose main task is to ensure
interoperability between cross-border traffic.

\ac{FRMCS} supports four levels of \ac{GoA} (numbered as 0-3), including automatic train protection, driver
advisory systems, \ac{ATO}, and driverless and unattended train operations.
Since \ac{FRMCS} supports \ac{ATO} in both urban and cross-country rail lines, the \ac{ETCS}
monitors the train's movement to ensure it adheres to the local speed limit
and its own permitted top speed \gf{and also ensures} that the train does not exceed
its operating authority (\gf{that is,} the location at which the train is permitted to travel in
a certain time window).
\PH{In addition, the \ac{ETCS} monitors track selectivity, train orientation, and direction of movement.
For some \ac{FRMCS} \ac{C-ITS} services, accurate absolute position of the train is required \cite{Talvitie:19}.}
At level 3 \ac{GoA}, \PH{for} train location and train integrity (\gf{that is,} the completeness of the train) supervision,
trains will rely on onboard sensors, \ac{GNSS} and cellular positioning \cite{Talvitie:19}.

The absolute longitudinal positioning accuracy requirements for \ac{ETCS} depend
on the specific \ac{FRMCS} and \ac{ERTMS} applications, involving accuracies down to 10 m \cite{STARS:17}.
For latitudinal accuracy -- \gf{that is cross rail track} -- positioning errors less than 2 m are required
\gf{in order to accurately determine} which track that is currently used by a particular train set.
Even stricter requirements on the onboard positioning system are imposed by higher levels of \acp{ATO},
\gf{that is}, driverless or unattended trains (see Table \ref{tab:tab1}).
For such operations, 
high accuracy positioning and situational awareness are required, especially when the train is in a station area.
The required accuracy can then be in the order of a meter down to a few decimeters.
At the same time, as meeting the accuracy requirements, the integrity level of the position must be \gf{high}
in order to meet the functional safety requirements of the specific applications. 

\subsection{Summary of Main Challenges for Positioning Algorithms}
In light of the \ac{V2X} and rail transport use cases and requirements, there are several challenges for positioning algorithms and supporting architecture.
High-accuracy radio-based positioning with low latency and strict requirements on integrity is difficult, even in favorable radio conditions, \gf{such as} when the vehicle is in the \ac{LoS} of its serving \ac{BS}. Instead, the positioning solutions need to exploit multiple input data, \gf{such as} onboard sensors and cellular measurements, which is a non-trivial problem. Both the vehicle's actual geographical position and the cellular measurements may evolve smoothly in time or undergo abrupt changes, which makes data selection and filtering challenging.
Moreover, if the vehicle is in \ac{NLoS} of its serving base station, deriving geometric information from the received radio signals is even more challenging.
However, recent research contributions indicate that multipath radio signals can also be used for determining position \cite{Monfared'2021}.

The decision on whether location-related computation  should  be  executed  in  the  vehicle or in the network is also non-trivial.
In many cases, the device has the \gf{capability} to implement the required positioning solution.
\gf{However,} for low-cost, or low-power devices, network-based positioning using uploaded sensor data is foreseen.
Such \gf{solutions} put new requirements on the network, and the positioning architecture, driving processing towards the edge nodes.
Additional architectural challenges are \gf{presented} by high mobility since \gf{device-related}
data (with edge-near processing) needs to follow the vehicle as it moves through the network.

\section{Positioning Support in 3GPP 5G New Radio Networks} 
\label{Sec:5GEvo}
Even though positioning services have been part of previous cellular generations, \ac{5G} allows for significant improvements. {It} supports much higher frequencies (up to 100 GHz), larger bandwidths, \gf{and} improved
	positioning capability using a positioning reference signal. Moreover, the sidelink in \ac{5G} has \gf{a} physical layer support
	for unicasting, which facilitates cooperative vehicle positioning \cite{Blasco'2020}. Due to these features, positioning in \ac{5G} can be
	downlink-based, uplink-based, sidelink-based \gf{or
	based} on a combination of these schemes.

Currently, the following key positioning methods, which are applicable either in \ac{UEA} and/or \ac{UEB} mode,
are supported by \ac{5G} systems \PH{as of \ac{3GPP} Rel-17}:
\begin{itemize}
\item
\textbf{\ac{DL-TDOA}} is based on device {\ac{TOA}} measurements, reported relative to a reference \ac{TOA} measurement (UEA, UEB).
\item
\textbf{\ac{DL-AOD}} is based on device downlink antenna beam measurements to estimate the elevation and azimuth angles relative the transmitting antenna (UEA, UEB).
\item
\textbf{\ac{UL-TDOA}} is based on network \ac{TOA} measurements (UEA).
\item
\textbf{\ac{UL-AOA}} \PH{exploits} 
multiple antenna elements 
to estimate the elevation and azimuth angles relative the device (UEA).
\item
\textbf{\ac{multi-RTT}} is based on a combination of downlink and uplink \ac{TOA} measurements relative
to a transmission time reference, which combine to a roundtrip time measurements to one or more \gf{transmission and reception points} (UEA).
\item
\textbf{\ac{GNSS} \ac{RTK}} is based on scalable and interoperable assistance data with corrections to enable high accuracy (UEB).
\item
\textbf{Hybrid positioning methods based on sensor measurements} \fg{were introduced already in Rel-15, by means of \ac{UE} {providing} movement information.
The movement information may contain displacement results, estimated as an ordered series of points.
This motion-sensor based positioning method can be combined with other positioning methods,
to facilitate hybrid positioning methods.
}
\end{itemize}

For accurate vehicular positioning, angular information plays an important role in many scenarios.
In \ac{5G}, the angle based \ac{DL-AOD}
positioning method is based on downlink timing measurements
of a \gf{\ac{DL-PRS}}, configured per resource,
where resources are combined into sets, which are associated with a \gf{transmission and reception point}.
In \ac{5G} \ac{FR2} -- that is, in mmWave bands -- a \ac{DL-PRS} resource is generally associated with a beamformed transmission.
With the large antenna arrays and dense deployments in these bands,
rich beam-based angular measurements can be provided.

The \gf{ongoing releases} (3GPP Rel-17 and Rel-18) are further addressing high accuracy,
positioning integrity, and sidelink positioning
with the features \gf{that are} most relevant for the transport sector \gf{described briefly} below. 

\begin{itemize}
\item
\textbf{High Accuracy:}
\PH{Support for \ac{LoS}/\ac{NLoS} detection and indication,} \ac{UL-AOA} and \ac{DL-AOD} enhancements 
\PH{(related to provisioning of assistance data to improve/simplify angle estimation), are part of Rel-17.}
Additionally, cellular carrier phase positioning, along with bandwidth aggregation for
intra-band carriers, as a means to increase the effective bandwidth and delay resolution will be studied in Rel-18.

\item
\textbf{Positioning integrity:}
\PH{Provisioning of GNSS integrity information is part of Rel-17. Integrity information for cellular positioning methods will be covered in Rel-18.}
In the integrity procedures, the network and device exchange information about anticipated events that may compromise positioning.

\item

\textbf{Sidelink positioning:}
Sidelink ranging and positioning in different coverage scenarios,
including out-of-coverage, will be studied in Rel-18.
\end{itemize}

\section{Architecture and Protocol Outlook for Location-Aware Services} 
\label{Sec:Arch}

\ph{In order to support signal acquisition from multiple sources
and to facilitate sensor fusion, which will be discussed in Section \ref{Sec:Perf}, architecture enhancements are required. This section discusses requirements and architecture solutions applicable
in \ac{5G} networks supporting vehicular use cases.}

\subsection{Architecture Requirements to Support Road and Rail Use Cases}
\label{Sec:ArchReq}

In \ac{5G}, the architecture and protocols support the provisioning of vehicle-based measurements to the network, more specifically to the \ac{LMF}. {The \ac{LMF} in the current 3GPP architecture is a central \ac{LMF}, which runs in a massive cloud platform and can be co-located with other core network entities, such as the \ac{AMF} \cite{ETSI:2018}.}
Examples of vehicle-based measurements are displacement readings from \ac{IMU} sensors,
and barometer pressure sensors for altitude computation and reporting.
Such information can be used to perform hybrid positioning at the network side,
which \gf{uses} other measurements or absolute positioning methods.
This enables the \ac{LMF} to exploit assumptions on device mobility and to achieve positioning enhancements through tracking.
\ph{In automotive and rail transport scenarios, many other sensors, such as
light sensors, radars, cameras, and lidar sensors are typically available on
the device (vehicle) side,
which \gf{provides} valuable information for vehicle positioning
and situational awareness.}
With appropriate protocol enhancements,
such information can be made available to the network, allowing for enhanced hybrid positioning solutions.

Such solutions may be relevant in scenarios, in which
the computational complexity may be prohibiting at the device side, (for example, in the case of low-cost vehicles like bicycles).
Furthermore, for safety critical applications, the network may need to validate
the position derived and reported by the device.
Future positioning computation engines must
process a vast amount of sensory and cellular measurement-based data.
In many cases, sensor information can be advantageously fused with positioning-related cellular measurements,
such as \PH{\ac{TOA} and \ac{AOA}.} 
Thus, the architecture must also address {in which entity}
the sensor fusion should take place in order to estimate and track the location of vehicles.
Finally, the architecture must also allow for fast access and exchange
of the rich set of information from sensors and provision for low-latency processing of data, including
computing position estimates.
To summarize, \gf{the fundamental building blocks of future network architectures are}
providing high-capacity storage, fast processing and
location information to clients by meeting the quality of service requirements in terms of latency
and accuracy.

\begin{figure}[t]
\begin{center}
\includegraphics[width=1\columnwidth]{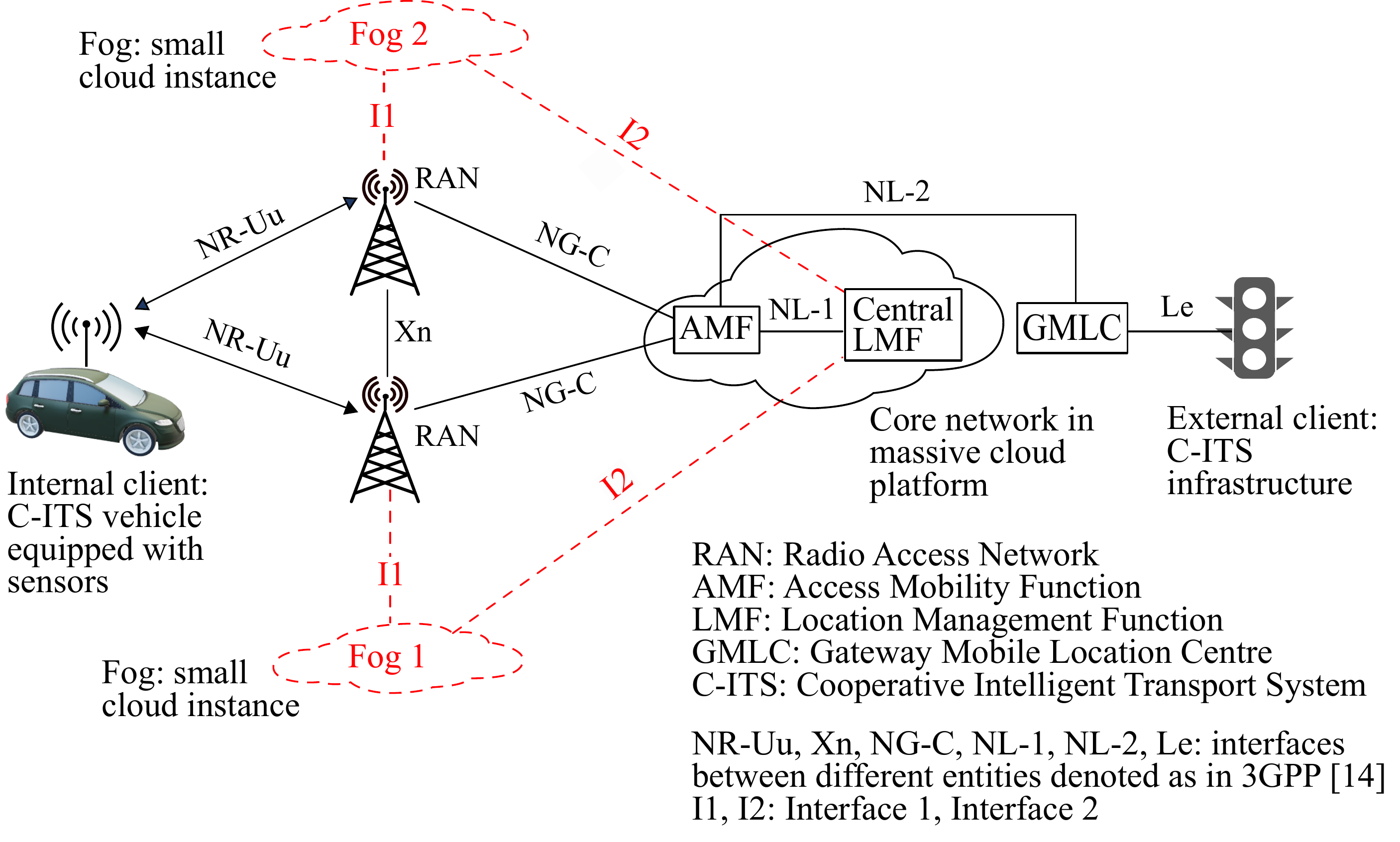}
\caption{{Enhanced architecture for sensor fusion for \ac{C-ITS}, with differences as compared to the current 3GPP architecture depicted by dashed red lines. The red parts represent the new proposed interfaces and fogs.}}
\label{FigAMF}
\end{center}
\end{figure}

\subsection{Proposed Architecture for Fusing Sensor and Cellular Measurements}

In the current positioning architecture \cite{TS:23.273}, a signal must pass through multiple hops before reaching the location server (\ac{UE}-\ac{BS}-\ac{AMF}-\ac{LMF}).
A decentralized architecture with a reduced number of hops can allow sensor fusion \PH{to} be performed with a reduced latency. 
Figure~\ref{FigAMF} illustrates an architectural solution in which the additions
to the current architecture are depicted \gf{by a dashed red line}.

{In the deployment scenario of Figure~\ref{FigAMF}, two different clients (receivers of the estimated position) are depicted; one network internal and one external \cite{TS:23.273}.
The internal client resides in the  \ac{C-ITS} vehicle, which is equipped with sensors, and seeks positioning information from the \ac{LMF}.
An external client -- such as one that is responsible for coordinating traffic and interaction with the fixed infrastructure -- is also required to determine the location of the vehicle.}
As indicated in Figure~1, in the proposed architecture, fog instances (denoted by Fog~1 and Fog~2) are defined and are located closer to the device in order to support \ac{MEC}. The fog instance of a given \ac{LMF} manages the location context of the device as long as the device is located within the cells belonging to the \ac{RAN} nodes managed by this fog instance. When the device leaves the area managed by the current fog instance (for example when transitioning from Fog~1 to Fog~2 as shown in Figure~\ref{FigAMF}), the location context of the device is sent to the central LMF, which may then forward it to another fog instance. In addition to fog instances, two new interfaces are defined, as shown in Figure~\ref{FigAMF}. The I1 interface is defined between the \ac{BS} and the fog instance, while the I2 interface is defined between the fog instance and the central \ac{LMF} cloud instance. The I1 interface \gf{makes it possible} to transfer UE positioning measurement reports and sensor measurement reports to the fog instance via the serving \ac{BS}. Similarly, the \ac{BS} measurements (including serving and non-serving \ac{BS} measurements) can also be provided to this fog instance. Hence, in actual deployments there may be several I1 interface instances between a fog instance and the (serving and non-serving) \acp{BS}.
The I2 interface \gf{makes it possible} to transfer \gf{the} computed position of the device to {the} central \ac{AMF} and \ac{LMF}.

The proposed fog instances and interfaces \gf{enable sensor fusion to occur} significantly closer to the UE with a reduced number of hops needed (UE-BS-fog).
	\gf{This allows to} significantly reduce latency and improve capacity storage \gf{compared}
	to available solutions using the current architectures.
\PH{Moreover, the fog is an entity belonging to a core
	network node, which is more secure compared to being in a RAN
	node since the UE identifier is only known to the core network
	nodes and not to the RAN nodes, and thus privacy and security
	can be preserved.
}

\section{Case Study: Kalman Filter-based Information Fusion} 
\label{Sec:Perf}

\subsection{Sensor Fusion to Enhance \ac{5G} Positioning Capabilities} 
\label{Sec:Sensor}

As mentioned previously, modern vehicles rely on a large set of sensors
and data sources {allowing} to acquire location awareness and positioning.
However, depending on the situation and conditions such as weather and visibility, some sensors may fail. On the top of this, \ac{GNSS} which is generally used as an absolute positioning source, potentially enhanced by using assistance data, may not be available in occlusion scenes such as tunnels and urban canyons.
Motivated by this observation, we propose and analyze a framework that is
suitable in \ac{GNSS} challenging environments. This framework
utilizes measurements by an onboard \PH{\ac{IMU} sensor} and
fuses these with \ac{5G} measurements, such as \gf{those} presented in Section~\ref{Sec:5GEvo}.
The \ac{IMU} {measures} the speed, acceleration, and orientation of the vehicle for position tracking. Such a framework can
be implemented at either the vehicle side, or the network
side by exchanging data over the standardized interfaces of
\ac{5G} as discussed in Section~\ref{Sec:Arch}.

It is worth noting that fusing
data provided by onboard sensors with measurements on radio
signals incurs some computational complexity, depending on
{the frequency of measurement updates; the amount of data provided by
each measurement and the algorithm(s) used to fuse such
measurement data.}

\subsection{Kalman Filter-based Information Fusion}
Due to its ability to track autoregressive processes, we propose a discrete Kalman filter framework suitably tailored
to the \ac{5G} architecture in order to track a vehicle's position at discrete time instances.
The Kalman filter approach requires the definition of a state transition equation and the associated measurement (observation) equation.
{The choice of the measurement sources} is a challenging design problem, which must 
be further addressed in future studies, as the number of available measurement sources \gf{is likely to increase} in the future.
\PH{For the considered vehicular tracking problem, the measurement sources consist of \ac{IMU} sensor data and cellular measurements.
To achieve high positioning accuracy, high-quality measurement data are essential, which, in the case of cellular measurements, are 
impacted by the deployment and propagation environment.} An associated high computational complexity may be acceptable when the Kalman filtering is running at the network side and takes advantage of \ac{MEC} resources, as discussed in Section~IV. However, when the computations are done at the \ac{UE}, the number of measurement sources in general must be limited for complexity reasons.
To aid signal source selection, future high-definition maps can include information
about the propagation conditions, availability, and quality of wireless signals.

\begin{figure}[h]
	\begin{center}
		
		\includegraphics[width=0.85\columnwidth]{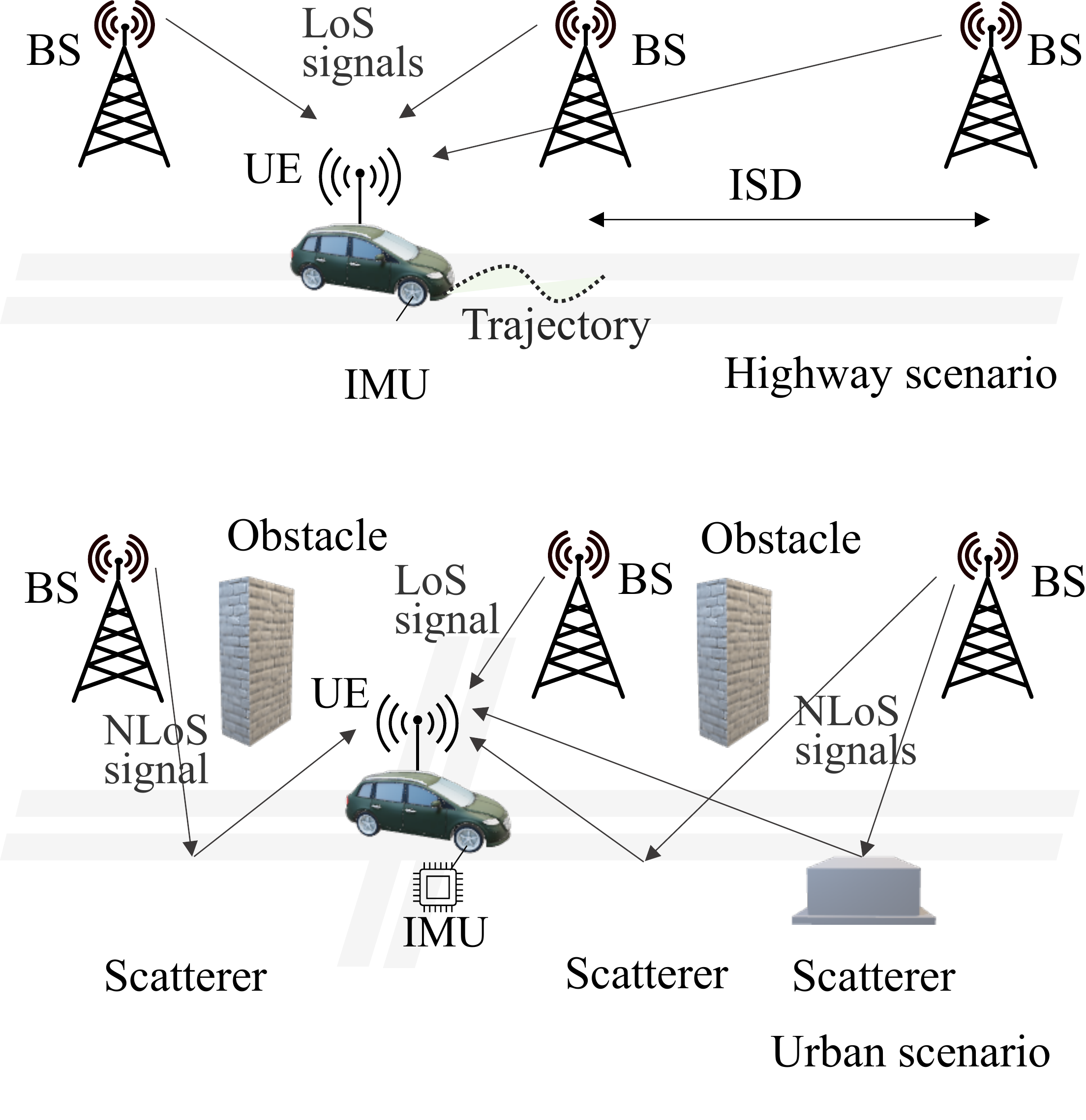}
		\vspace{0.1cm}
		\caption{{High-way scenario with \ac{LoS} signals from multiple \acp{BS} (top figure) and dense urban scenario with a mix of \ac{LoS} and \ac{NLoS} signals from different \acp{BS} (bottom figure).}}
		\label{Fig:Rural_vs_urban}
	\end{center}
\end{figure}

{In this case study, we consider a vehicle equipped with an onboard motion sensor and moving in a highway scenario as illustrated in the top part of Figure~\ref{Fig:Rural_vs_urban}.} In this scenario, the \ac{LoS} channel propagation is typical, and measurement signals from multiple \acp{BS}
(typically from several closest \acp{BS} as depicted in the top part of Figure~\ref{Fig:Rural_vs_urban})
are available at the vehicle. Hence, these can be used in the Kalman filter as sources of input data.

\begin{figure}[t]
	\begin{center}
		
		\includegraphics[width=1\columnwidth]{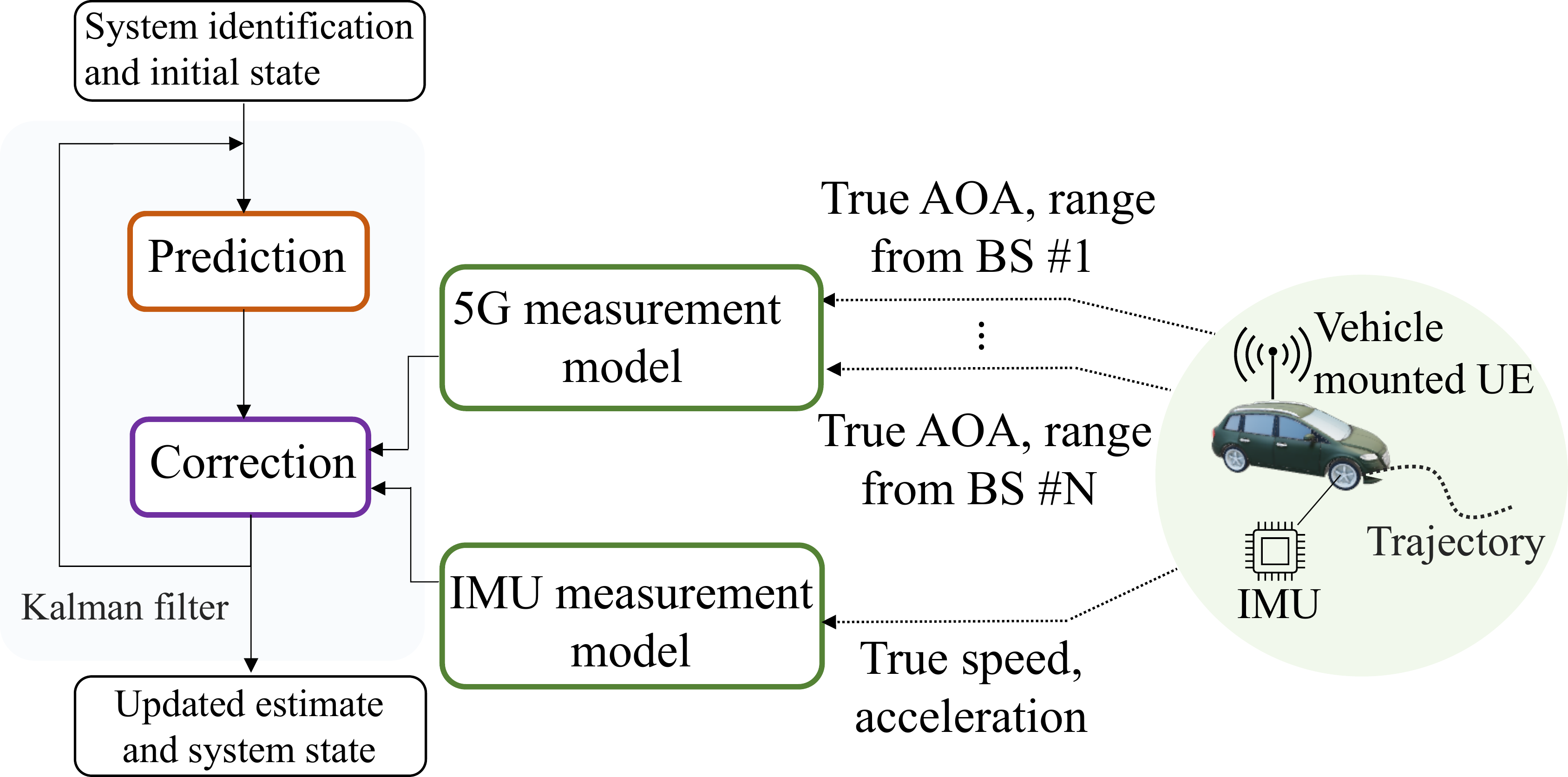}
				\vspace{0.2cm}
		\caption{{Flowchart of an extended Kalman filter-based position estimation method using model-based AOA and range measurements from $N$ \acp{BS}.}}
		\label{Fig:FlowChart}
	\end{center}
\end{figure}

{The flowchart of the proposed position tracking algorithm based on extended Kalman filter is depicted in Figure~\ref{Fig:FlowChart},
which corresponds to the one used for the simulation results. The system state includes the speed, the acceleration, and the position of the vehicle in $(x,y)$ coordinate plane. The initial position is calculated based on the initial guess which can be obtained from a \ac{GNSS} signal.
Based on the system's state transition matrix and the updated system state, the Kalman filter scheme makes
a new prediction of the subsequent state. This predicted state is corrected based on the newly
incoming measurement data (speed and acceleration from the \ac{IMU} sensor and \ac{AOA} and range measurements from $N$ \ac{BS}s) to yield the next updated system state.}

\subsection{Simulation Results}
\label{Sec:SimRes}
\begin{figure}[t]
	\begin{center}
		\includegraphics[width=0.95\columnwidth]{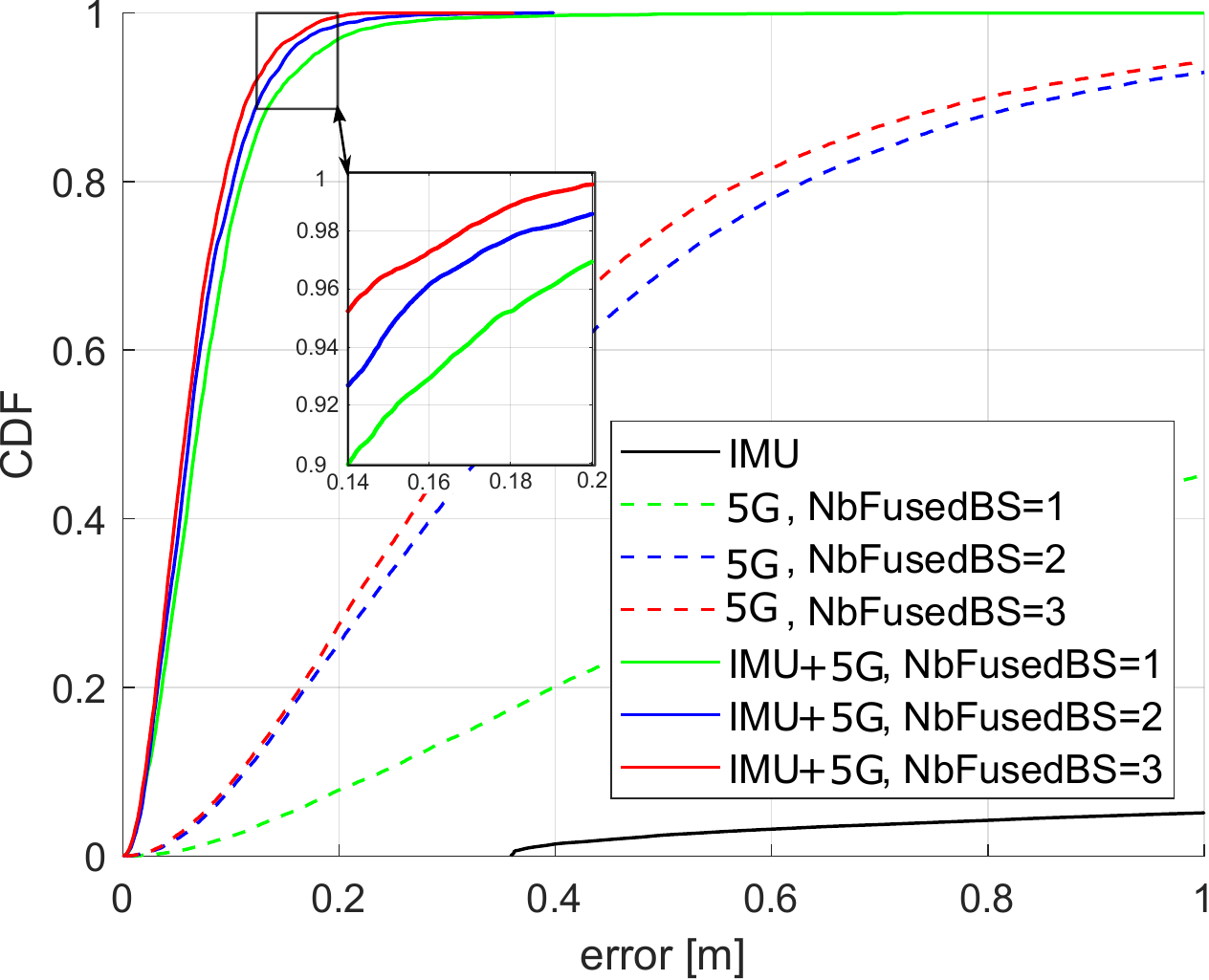}
		\vspace{0.2cm}
		\caption{CDF of the obtained positioning accuracy when using {IMU only},
			5G only and fused IMU+5G positioning for ISD=200 m.}
		\label{FigFus2}
	\end{center}
\end{figure}

\begin{figure}[t]
	\begin{center}
		\includegraphics[width=0.95\columnwidth]{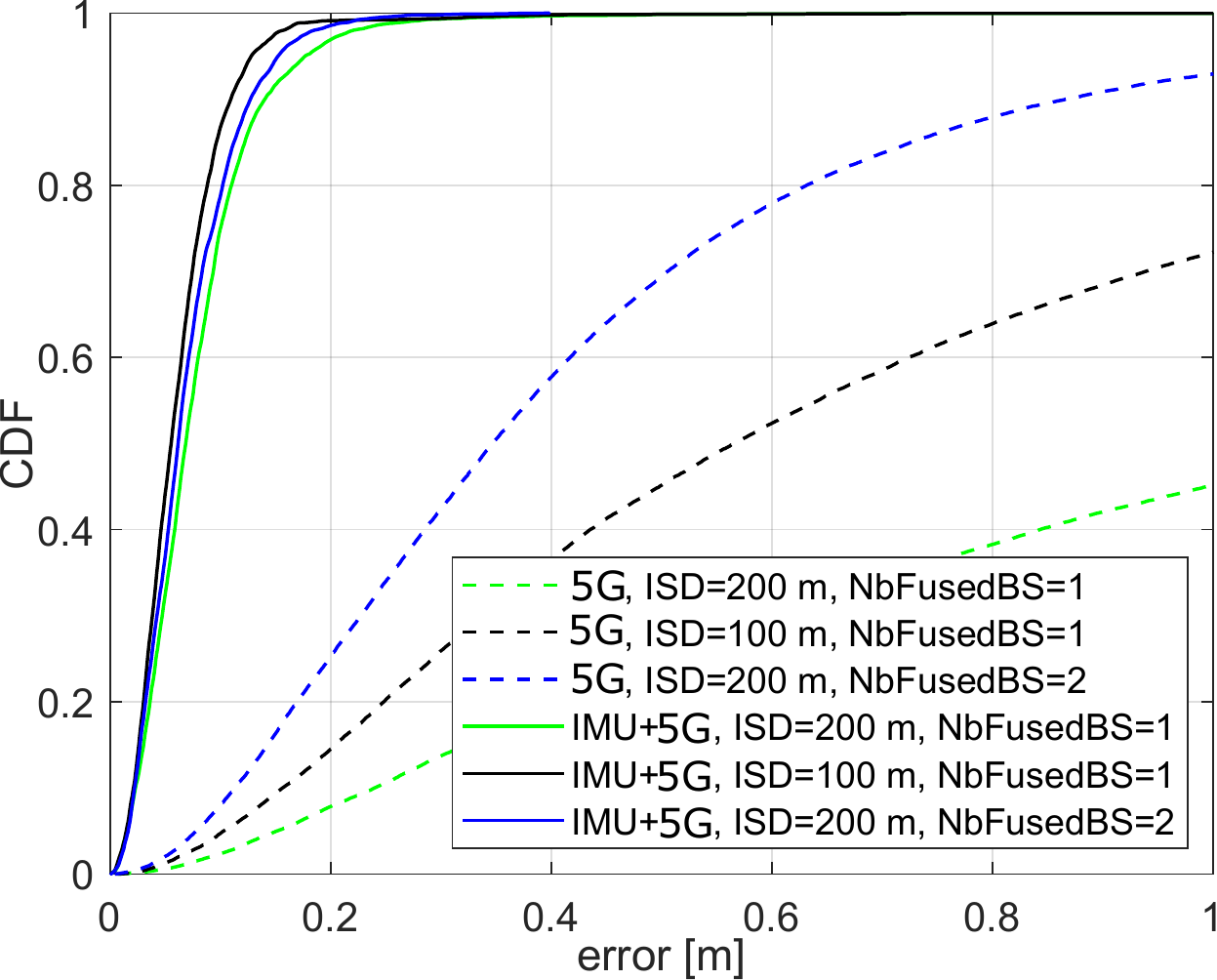}
				\vspace{0.2cm}
		\caption{CDF of the obtained positioning accuracy for different ISDs and numbers of fused base stations.}
		\label{FigFus3}
	\end{center}
\end{figure}

We consider a highway scenario with equally spaced \acp{BS}
placed 30 meters from the road. A moving vehicle with a speed of 130 km/h is equipped with an onboard \ac{IMU} sensor and an \ac{5G} \ac{UE}.
A \ac{MIMO} system is considered with square antenna arrays at the \ac{BS} with 256 antennas and the \ac{UE} with four antennas.
A \ac{LoS} downlink propagation scenario is assumed with a grid of Discrete Fourier Transform beams transmitted
by the \ac{BS}s.
The operating frequency is assumed to be 28 GHz with a transmit power of 40 dBm.
The update rate of the \ac{IMU} is equal to the downlink signal periodicity, which is assumed to be 100ms. The total travelled distance over which the results are averaged is equal to 10km.
The \ac{5G} measurements consist of range and \ac{AOA} measurements and an analytical error model is used to generate the individual samples \cite{Mostafavi:20}.
Similarly, the noisy acceleration data is generated based on an \ac{IMU} measurement model \cite{Mostafavi:20}.

Based on the Kalman filter presented in Figure~\ref{Fig:FlowChart}, we compare three different fusion-based positioning approaches:
(1) \ac{5G} combined with IMU: a sensor fusion-based method using \ac{IMU} measurements
and \ac{5G} downlink measurements from $N$ \ac{BS}s;
(2) \ac{5G} only: a method based on \ac{5G} downlink measurements from $N$ \ac{BS}s without \ac{IMU} measurements;
(3) \ac{IMU} only: an \ac{IMU}-based positioning method without radio-measurements.

The positioning accuracies in terms of their \acp{CDF} are compared in Figure~\ref{FigFus2}.
The \ac{ISD}, which is the distance between two neighboring \ac{BS}s, is assumed to be equal to 200 m, while the number of fused \ac{5G} measurements,
denoted by {NbFusedBS=$N$, varies from 1 (single \ac{BS}) to 3.}
The simulation results show that a large performance gain is obtained for the sensor fusion-based method compared to the \ac{5G}-only-based method.
{A particularly poor performance is obtained for the IMU-only based method which is explained by the accumulation of positioning errors over the total travelled distance}.
\gf{Note} that fusion-based methods are able to achieve a decimeter level accuracy
with greater than 90\% probability.
However, the \ac{5G} only-based method can provide a sub-meter
accuracy when fusing measurements from multiple \acp{BS}.

Figure~\ref{FigFus3} emphasizes the importance of fusing measurements from multiple \acp{BS}.
This is especially important for the \ac{5G} only-based method, for which an important gain is achieved
by fusing measurements from 2 \acp{BS} for the ISD=200 m compared to a two times more densified scenario with an \ac{ISD} of 100 m. \jv{Hence, exploiting measurements from multiple BSs is more beneficial than densification.}
However, when \ac{5G} measurements are combined with \ac{IMU}, fusing measurements from multiple \acp{BS}
\jv{provides only a small gain in positioning accuracy.}

 \subsection{Discussion and Future Work}

The simulation results provided in the previous section are restricted to \PH{a} LoS scenario, which in general simplifies the positioning problem and channel acquisition process. \gf{The bottom part of Figure \ref{Fig:Rural_vs_urban} depicts a different scenario}, including \ac{LoS} and \ac{NLoS} regions, which is typical in an urban environment. The figure illustrates some of the challenges faced in such scenario, where the wireless channel experiences abrupt changes as the vehicle moves along a road. Additionally, \ac{NLoS} channel measurements pose both challenges and opportunities for positioning which need to be considered in future work.

It should be noted that the achievable positioning accuracy is highly dependent of the quality of the channel state information \PH{obtained for the cellular signals}. This is especially true when multipath propagation is exploited through spatial measurements. Therefore, in addition to the vehicle's position tracking, channel tracking is essential for achieving a high positioning accuracy. To this end, \gf{future work could consider} a two-stage Kalman filter for tracking both the vehicle's position and the wireless channel.
One advantage of separating channel and position tracking is that channel estimates can be exposed
to other functions running at the \ac{UE}, \gf{such as} data demodulation and decoding.

An additional challenge is the increased complexity of the positioning problem when exploiting multi-path propagation.
This speaks in favor of \ac{MEC}-based positioning and tracking, where computational complexity is less \gf{prohibitive}.
This may call for additional measurements being standardized and exchanged over the 3GPP interfaces in the future.

\section{Concluding Remarks and Outlook}
\label{Sec:Conc}
As the automotive and rail industries and surrounding ecosystems define and experiment with new use cases,
there is a growing interest in high-accuracy localization services that determine and predict vehicle positions
in high-speed and high-vehicle-density environments.
\gf{To meet the increasing expectations by the automotive and rail industries,
recent advances in using cellular signals and measurements to determine the position of connected cars,
trains and vulnerable road users provide technology enablers for transport applications.}
We \gf{have} argued that fusing \ac{IMU} measurements with cellular signals is highly non-trivial from both {the signal processing and
the architecture and protocol point of view.} Our results indicate that combining measurements from multiple
\acp{BS} and taking advantage of locally available sensor measurements can meet stringent localization
requirements under proper deployment of the cellular infrastructure.

\gf{Further improving} the performance and reliability of real-time localization algorithms that take advantage of multipath signals,
multiple base stations and various sensor measurements requires further research.
An important open research question concerns the distribution of localization functionalities between vehicles and networks nodes, which may have
far-reaching consequences on the inherent trade-offs \gf{among} localization accuracy, reliability, latency, and radio interface
resources required for the communication between mobile and infrastructure nodes.

\bibliographystyle{IEEEtran}
\bibliography{D2D5Gv10,IEEEfull}

\vspace{0.5cm}

\footnotesize{
\small{
\noindent \textbf{Peter Hammarberg} is a senior researcher at Ericsson, working in the area of wireless communication and radio localization.
He holds a Ph.D. in radio systems from Lund University, Sweden.

\noindent \textbf{Julia Vinogradova} received her Ph.D. degree in electrical engineering from Telecom Paris-Tech, France, in 2014. From 2015 to 2016, she was with the Communication Systems Division at Linköping University, Sweden. Currently, she is with Ericsson Research, Jorvas, Finland. 

\noindent \textbf{Gabor Fodor} is a master researcher with Ericsson Research and an adjunct professor with
the KTH Royal Institute of Technology. He holds a Ph.D. degree from the Budapest University
of Technology and a docent degree from KTH. He is an Editor of IEEE Wireless Communications and
IEEE Transactions on Wireless Communications.

\noindent \textbf{Ritesh Shreevastav} is a senior researcher and 3GPP RAN2 delegate in Ericsson Research.
He received the M.Sc. degree in Telecommunications from Queens University Belfast, UK (2006)
and a Research Masters in Computer Science (Licentiate) from Trinity College Dublin, Ireland (2008).

\noindent \textbf{Satyam Dwivedi} is a senior researcher at Ericsson research and team leader for positioning
research and standardization team. He holds an Ms.C and a Ph.D.degree from the Indian Institute of Science, Bangalore, India.

\noindent \textbf{Fredrik Gunnarsson} is an expert in RAN automation and positioning at Ericsson Research.
He obtained his M.Sc. and Ph.D. in 1996 and 2000 respectively, both in Electrical Engineering from Link\"{o}ping University,
Sweden, where he also holds a position as adjunct professor.

}

\end{document}